\documentclass[%
 reprint,
 amsmath,amssymb,
 aps,
prd,
]{revtex4-2}
\usepackage{xspace}
\usepackage{graphicx}
\usepackage{dcolumn}
\usepackage{natbib}
\usepackage{bm}
\usepackage{xcolor}
\usepackage{lineno}

\usepackage[colorlinks=true,linkcolor=blue,citecolor=blue,urlcolor=blue]{hyperref}
\newcommand{\am}{\textsc{AM$^3$}\xspace}

\usepackage{fontspec}
\usepackage{xeCJK}

\begin{document}

\preprint{APS/123-QED}

\title{Coupled Time-Dependent Proton Acceleration and Leptonic-Hadronic Radiation in Turbulent Supermassive Black Hole Coronae}

\author{Chengchao Yuan (袁成超)$^1$}\email[Corresponding author: ]{chengchao.yuan@desy.de}
\author{Damiano F.\ G.\ Fiorillo$^1$}%
\author{Maria Petropoulou$^2$}
\author{Qinrui Liu (刘沁枘)$^{3,4,5}$}
 
\affiliation{$^1$Deutsches Elektronen-Synchrotron DESY,
Platanenallee 6, 15738 Zeuthen, Germany}%
\affiliation{$^2$Department of Physics, National and Kapodistrian University of Athens, University Campus Zografos, GR 15784, Athens, Greece }
\affiliation{$^3$Department of Physics, Engineering Physics and Astronomy, Queen's University, Kingston ON K7L 3N6, Canada}
\affiliation{$^4$Arthur B. McDonald Canadian Astroparticle Physics Research Institute,  Kingston ON K7L 3N6, Canada}
\affiliation{$^5$Perimeter Institute for Theoretical Physics, Waterloo ON N2L 2Y5, Canada}

\date{\today}
\begin{abstract}
{Turbulent coronae of supermassive black holes can accelerate non-thermal particles to high energies and produce observable radiation, but capturing this process is challenging due to comparable timescales of acceleration, cooling, and {the development of cascades}. We present a time-dependent numerical framework that self-consistently couples proton acceleration—modeled by the Fokker-Planck equation—with {leptonic-hadronic radiation}. For the neutrino-emitting Seyfert galaxy NGC~1068, we reproduce the neutrino spectrum observed by IceCube, while satisfying gamma-ray constraints. We also consider a transient corona scenario, potentially emerging in tidal disruption events like AT 2019dsg, and show that cascade feedback on proton cooling can impact proton acceleration and radiation processes in weaker coronae, producing delayed optical/ultraviolet, X-ray, and neutrino emissions of $\mathcal O(100~\rm d)$. This flexible tool efficiently models multi-messenger signals from both steady and transient astrophysical sources, providing insights in combining particle acceleration and radiation mechanisms.}

\end{abstract}
\maketitle
\section{Introduction}
The recent detection of high-energy neutrinos from the nearby Seyfert galaxy NGC 1068 by IceCube \citep{IceCube:2019cia,IceCube:2022der} provides compelling evidence for particle acceleration in active galactic nuclei (AGN). The neutrino spectrum, which peaks around 1–100 TeV, suggests a coronal origin \citep{Murase:2019vdl,Inoue:2019yfs, Kheirandish:2021wkm,Padovani:2024ibi} where protons could be accelerated by magnetized turbulence \citep{Murase:2019vdl, Fiorillo:2024akm,Saurenhaus:2025ysu} or magnetic reconnection \citep{Kheirandish:2021wkm,Mbarek:2023yeq,Fiorillo:2023dts,Karavola:2024uui} {in a hot and dense region, known as corona,} near the supermassive black hole (SMBH). These protons interact with ambient X-ray and optical/ultraviolet (OUV) photon fields to produce neutrinos and electromagnetic (EM) cascades {(initiated by secondary particles)} ~\citep{Murase:2019vdl,Kheirandish:2021wkm,Fang:2022trf,Fang:2023vdg}. {In this model, the gamma-rays could be efficiently absorbed due to $\gamma\gamma$ annihilation to satisfy the constraints \citep{Fermi-LAT:2019yla,MAGIC:2019fvw,Ajello:2023hkh}.} Moreover, IceCube has reported neutrino emission excesses from several additional X-ray–bright Seyfert galaxies, suggesting a disk–corona origin for the neutrinos \citep{IceCube:2024dou,IceCube:2023tts}. A self-consistent treatment is generally needed to incorporate EM cascade feedback {on proton cooling and acceleration processes} in the compact coronal regions. 

{Besides AGN coronae, which we consider to have achieved a steady state despite possible long-term variability \cite{Ursini:2015eda,2016AN....337..557W,Jana:2025pir,Laha:2024tnc}}, tidal disruption events (TDEs), in which the bound debris of destroyed stars powers a year-long transient spanning a broad electromagnetic spectrum \citep{Rees:1988bf}, offer a unique laboratory to study particle acceleration in evolving coronae likely formed during the super-Eddington accretion phase \citep{Murase:2020lnu}. Meanwhile, observed OUV and X-ray light curves (e.g., Refs. \cite{Auchettl:2016qfa,vanVelzen:2020cwu}) provide critical constraints. 

Modeling the {coronal emission, especially from transients,} generally requires coupling time-dependent proton acceleration with radiative feedback across multiple spatial and temporal scales. So we provide, for the first time, a systematic description that couples them together, within a Fokker-Planck realization of acceleration (e.g., Refs. \cite{1970JCoPh...6....1C,park1996stochastic,Becker:2006nz,Stawarz:2008sp}); we validate {our time-dependent code} by applying it to a steady environment that is crucial for multi-messenger astrophysics, namely the NGC 1068 corona, and then we prove its power by applying to a time-dependent setup for TDE transient coronae. Applying to NGC 1068 reproduces the observed neutrino spectrum while satisfying gamma-ray constraints, {as an expansion upon Ref. \cite{Fiorillo:2024akm} where cascade emission was not calculated.} For TDEs such as AT 2019dsg \citep{Stein:2020xhk}, we propose a phenomenological approach linking the time-dependent mass accretion rate onto the SMBH with transient corona properties, and demonstrate how early-stage cascade feedback shapes delayed multiwavelength and neutrino emission.

Another key motivation of this work is to provide the community with a versatile and efficient tool well suited for multi-messenger modeling of a wide range of astrophysical transients (e.g., gamma-ray bursts, AGN flares) and steady sources, accommodating diverse acceleration mechanisms such as turbulence (e.g., Refs. \cite{2012SSRv..173..535P,Xu:2017ypi}), magnetic reconnection (e.g., Refs. \cite{zweibel2009magnetic,Lazarian:2012nd,Guo:2014via,Guo:2015ydj,2025arXiv250602101S}), shearing flows (e.g., Refs. \cite{Rieger:2019uyp,Kimura:2017ubz}), and shock acceleration (e.g., Refs. \cite{Drury:1983zz,blandford1987shocks,Weidinger:2014yya}). 

{This paper is organized as follows. In \S\ref{sec:numerical}, we present a numerical framework that couples time-dependent proton acceleration with leptonic–hadronic radiation processes. We then apply this tool to model the multi-messenger emission from both steady (e.g., NGC 1068) and transient (e.g., non-jetted TDE) SMBH coronae in \S\ref{sec:applications}. In \S\ref{sec:discussion}, we discuss the corona model, the significance of radiation-feedback effects on the acceleration processes, and the flexibility of the framework in incorporating various particle-acceleration mechanisms, along with recent progress in this direction. A concluding summary is provided in \S\ref{sec:summary}.}

\section{Numerical Framework}\label{sec:numerical}

\begin{figure*}[htp]
    \centering
    \includegraphics[width=0.49\linewidth]{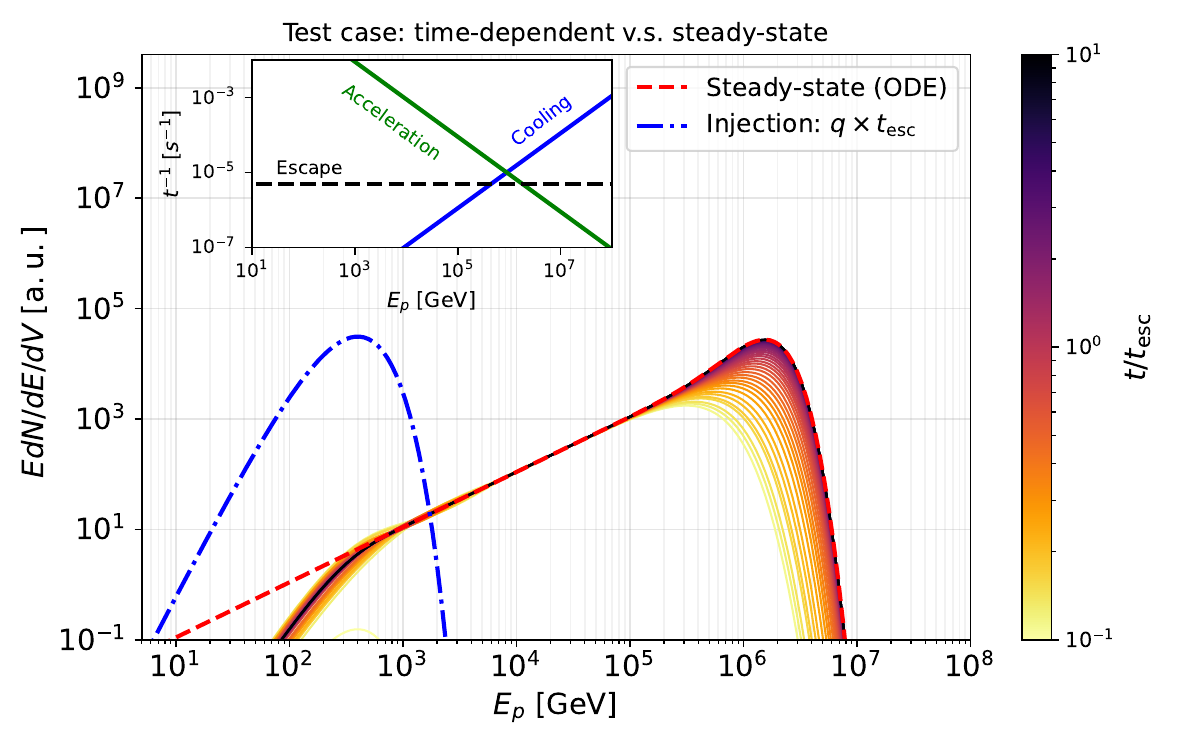}
    \includegraphics[width=0.49\linewidth]{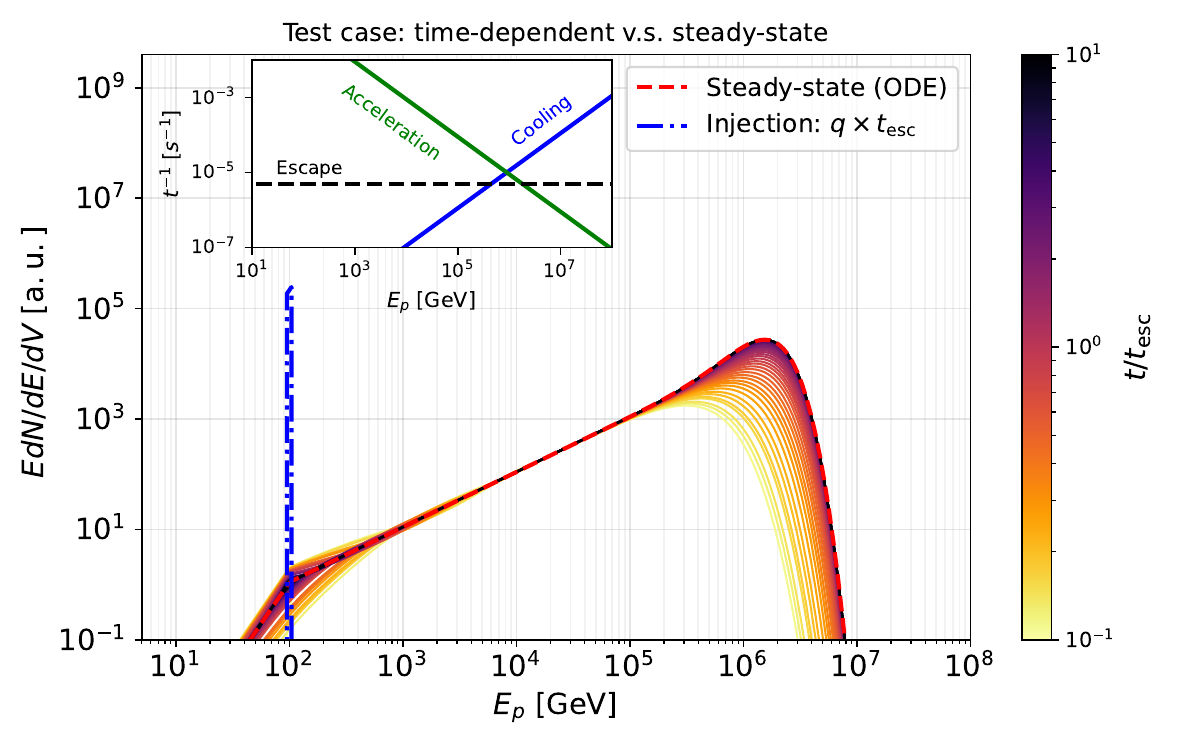}
    \caption{Test of numerical solutions to FP equations: comparison between time-dependent solutions (solid curves) and steady-state solutions (red dashed curve), normalized to arbitrary units. The injection rates (blue dash-dotted curve) are also shown. The insets display the proton escape, acceleration, and cooling rates. Different injection functions $q\propto p\exp(-p/p_{\rm inj})$ and $q\propto \delta(p-p_{\rm inj})$ with $p_{\rm inj}=10^2m_pc$ are used respectively in the left and right panels.}
    \label{fig:FP_test}
\end{figure*} 

Given the momentum diffusion term \( D_p(p) \) as a function of particle momentum $p$, the time-dependent Fokker-Planck (FP) equation can be used to describe proton acceleration in momentum space (e.g., Refs. \cite{1970JCoPh...6....1C,park1996stochastic,Becker:2006nz,Stawarz:2008sp}) incorporating cooling via lepto-hadronic radiation and escape from the acceleration zone,
\begin{equation}
    \frac{\partial f}{\partial t}=\frac{1}{p^2}\frac{\partial}{\partial p}\left[p^2D_{p}(p)\frac{\partial f}{\partial p}+\frac{p^3}{t_{\rm cool}} f\right] -\frac{f}{t_{\rm esc}}+q(p),
    \label{eq:FPT1}
\end{equation}
where $f(p,t)=\frac{dN_p}{dp^3dV}$ is the particle distribution, i.e., the number of protons per unit spatial volume $dV$ and per unit momentum-space volume $d^3p$,
$q(p)=\frac{dN_p}{dp^3dVdt}$ is the proton injection rate, $t_{\rm cool}$ is the proton cooling timescale, and $t_{\rm esc}$ is the proton escape time scale. The momentum diffusion term \( D_p \) is related to the acceleration timescale as \( D_p = p^{2} / t_{\rm acc} \). 

To solve the FP equation in a fully time-dependent manner, we adopt the Chang-Cooper method \cite{1970JCoPh...6....1C} to ensure both the preservation of positivity and the correct equilibrium solution, and use the Crank–Nicolson \cite{CrankNicolson1947} implicit time discretization scheme to rewrite the equation in a tri-diagonal matrix form. The system can efficiently evolve from  \( t \) to \( t+\Delta t \) by applying the inverse of a tridiagonal matrix \citep{code}, as described in Appendix \ref{app:numerical}. The implementation of the Chang-Cooper method and a reasonable time step $\Delta t$ significantly improve stability even for the cooling-dominant regime ($t_{\rm cool}^{-1}\gg t_{\rm acc}^{-1},~t_{\rm esc}^{-1}$). For a SMBH corona of radius $R_{\rm co}$, $\Delta t\sim0.001 R_{\rm co}/c$ would be sufficient. 

We then couple the FP equation with \am \cite{Klinger:2023zzv}, {an open-source software for time-dependent lepto-hadronic radiation modeling with optimized solvers}, which works with the proton density distribution $n_p(E,t)=Ed^2N_p/(dEdV)=4\pi p^3 f(p,t)|_{E=pc}$, we normalize $f(p,t)$ using the proton power $L_p$ and the volume of the acceleration zones $V$, 
\begin{equation}
{L_p}=-4\pi cV\int p^2D_{p}(p)\frac{\partial f}{\partial p}dp.
\label{eq:FP_norm}
\end{equation} One critical point is that, to avoid repeated proton cooling, we use \am solely to evolve the distributions of photons, neutrinos, and other secondary particles such as \( e^\pm \), \( \mu^\pm \), \( \pi^\pm \), and \( \pi^0 \) produced by \( p\gamma \), \( pp \), \( \gamma\gamma \), proton synchrotron (p-sy) and Bethe-Heitler (BH) interactions. The evolution of the proton spectra is determined exclusively by the FP equation, which already incorporates escape and cooling processes. Accordingly, after each time step, we update \( n_p(E,t) \) in \am, {and retrieve the overall, real-time cooling timescale from \am to the FP solver, e.g., \( t_{\rm cool} = \left( t_{pp}^{-1} + t_{p\gamma}^{-1} + t_{\rm BH}^{-1} + t_{\rm p\mbox{-}sy}^{-1} \right)^{-1} \), where \( pp \), \( p\gamma \), BH, and proton synchrotron cooling rates are included. Depending on the dynamics of the accelerator, the injection rate \( q \), acceleration time \( t_{\rm acc} \), and escape time \( t_{\rm esc} \) can take arbitrary forms and be time-dependent.}

To verify the robustness of the numerical methods presented above, we design a simple test case featuring free escape, shock acceleration with \( t_{\rm acc} \propto p \), and synchrotron cooling with \( t_{\rm cool} \propto p^{-1} \). We select a constant escape time and time-independent acceleration and cooling rates, given by
\[
t_{\rm esc} = 2 \times 10^5~\mathrm{s}, ~ t_{\rm acc} = 10^6 \left(\frac{p}{p_c}\right) \mathrm{s}, ~ t_{\rm cool} = 10^4 \left(\frac{p}{p_c}\right)^{-1} \mathrm{s},
\]
where \( p_c = 10^7 m_p c \) is a characteristic proton momentum. All other processes are switched off, and no feedback effects on the radiation field are considered. We also validate the convergence of the time-dependent solutions to a steady state solution by solving the steady-state FP equation by setting $\partial f/\partial t = 0$ (see Appendix \ref{app:numerical} for a detailed description of the numerical methods).

Fig. \ref{fig:FP_test} summarizes the results of the test case. The solid curves in color represent the proton distributions obtained from the time-dependent FP equation at times ranging from \( 0.1\, t_{\rm esc} \) to \( 10\, t_{\rm esc} \). The red dashed curve shows the steady-state solution. To verify the independence of $q$, the injection rates \( q(p) \propto p \exp(-p/p_{\rm inj}) \) and $q(p)\propto\delta(p-p_{\rm inj})$, with \( p_{\rm inj} = 10^2 m_p c \), are respectively used in the left and right panels. By comparing the time-dependent solutions with the steady-state solutions, we summarize the following conclusions:

\begin{itemize}
    \item The time-dependent solution obtained from solving the complete FP equation (partial differential equation) converges efficiently and stably to the steady-state solution derived from the reduced ordinary differential equation.
    \item The proton distribution for \( p > p_{\rm inj} \) is independent of the shape of the injection function \( q \) in the time-dependent treatment.
    \item The maximum energy of accelerated protons is consistent with the energy determined by the balance condition \( t_{\rm acc}^{-1} = t_{\rm cool}^{-1} \) (see the insets of Fig.~\ref{fig:FP_test}).
\end{itemize}

Regarding code efficiency, the time-dependent FP solver is fast, typically converging to a steady state within one minute if EM cascade feedback from \am is turned off. Most of the computational power is devoted to the leptonic-hadronic modeling by \am, especially in compact coronae, which requires very small time steps. Leveraging the efficient leptonic-hadronic modeling software \am, we find a way to overcome the challenge of reconciling the tiny time steps in compact acceleration regions with long-term system evolution. 

\section{Applications to SMBH coronae}\label{sec:applications}
\begin{figure*}[htp]\centering
       \includegraphics[width = 0.49\textwidth]{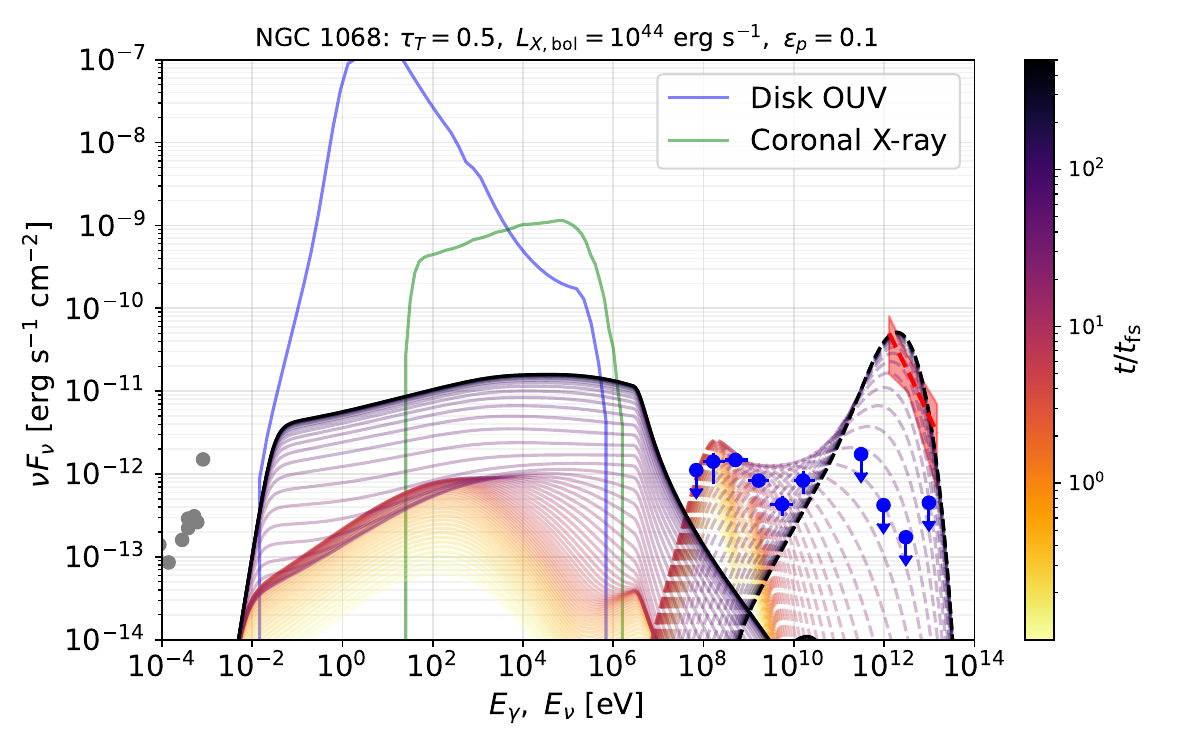}
    \includegraphics[width = 0.49\textwidth]{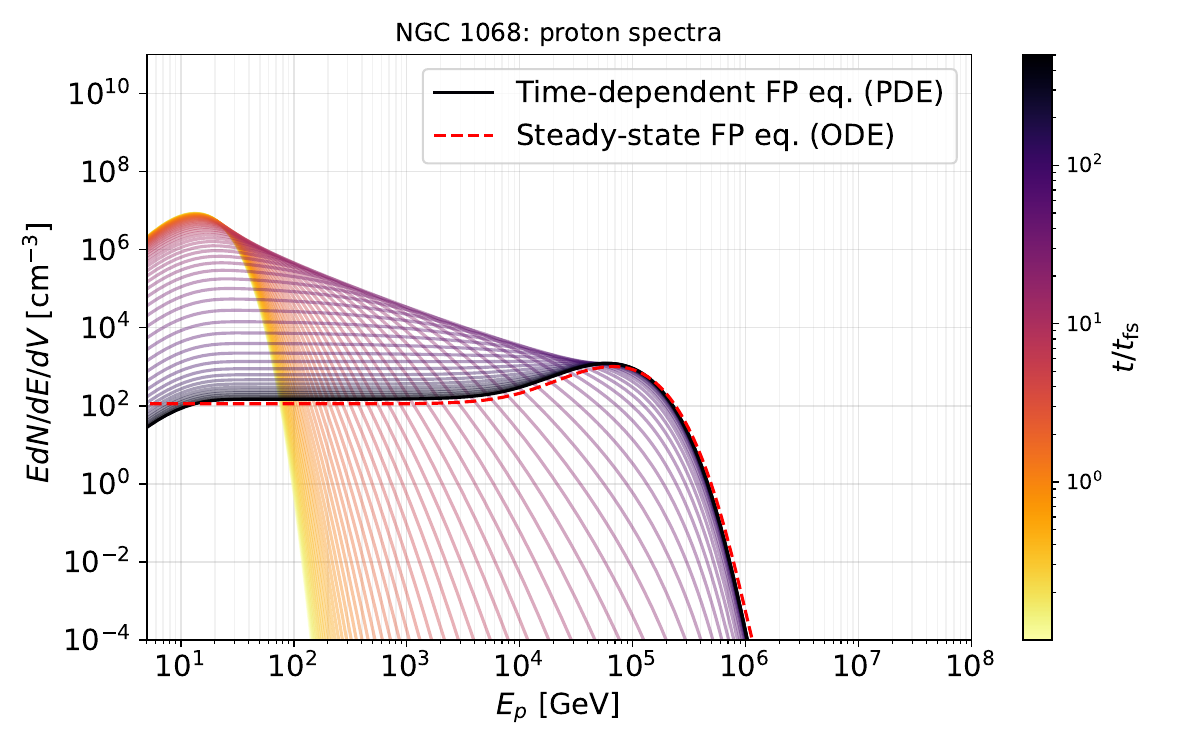}
          \caption{NGC~1068: Observed all-flavor neutrino (dashed) and EM cascade (solid) spectra (left panel), and in-source proton density spectra (right panel) at times ranging from $0.1t_{\rm fs}=0.1R_{\rm co}/c$ to $500t_{\rm fs}$. Radio \cite{Chang:2019cdu}, gamma-ray \cite{Fermi-LAT:2019yla,MAGIC:2019fvw}, and neutrino \cite{IceCube:2022der} observations are shown as gray points, blue points, and the red-shaded area. The green and blue curves respectively show the coronal X-ray and OUV spectra. The solid curves in the right panel show the proton spectra from time dependent FP equation, whereas the dashed red curve denotes the steady-state solution from the reduced ordinary differential equation (ODE). 
          }
          \label{fig:NGC1068}
\end{figure*}

{This framework's primary range of applicability includes SMBH coronae; crucial cases include the multi-messenger observations of NGC 1068 and non-jetted TDEs. We first consider proton acceleration by coronal turbulence, then benchmark our code against the stationary case of NGC 1068, and finally discuss its applicability to time-dependent setups in the context of TDEs.}

\subsection{Turbulent particle acceleration in coronae}\label{sec:acceleration}
We consider proton acceleration by magnetized turbulence within a corona, which refers to a spherical region of radius $R_{\rm co}$ around a SMBH, permeated by a magnetic field of strength $B$, and containing thermal protons and electrons with densities $n_p$ and $n_e$, respectively. Considering an SMBH of mass $M = 10^7 M_7 M_\odot$, we define the radius parameter $\mathcal{R} = R_{\rm co} / r_{\rm g}$ in terms of the gravitational radius $r_{\rm g} = GM / c^2 \simeq 1.5 \times 10^{12} M_7~{\rm cm}.$ Given the typical opacity $\tau_T \sim0.1-1$ \cite{Ricci:2018eir}, we infer the electron density $n_e = \tau_T / (\sigma_T R_{\rm co})$. For a steady corona in the core of an AGN, such as NGC 1068, we fix $\tau_T = 0.5$ and assume {$n_p \approx n_e =\tau_T / (\sigma_T R_{\rm co})\simeq 2.5 \times 10^{10} M_7^{-1} (\mathcal R/20)^{-1}~{\rm cm}^{-3}$} (as in Refs. \cite{Murase:2019vdl,Fiorillo:2024akm}). Given the magnetization $\sigma_B$ of the plasma, the magnetic field strength can be parameterized as
$B = \sqrt{4 \pi \sigma_B n_p m_p c^2}$, where $m_p$ is the proton mass.

Inside magnetized coronae, particles can be accelerated via plasma turbulence or magnetic reconnection. In this work, we focus on the turbulent acceleration scenario with $\mathcal R = 20$ and $\sigma_B = 0.1$. The strength of the turbulent magnetic field fluctuations can be written as $\sigma_{\rm tur} = \sigma_B (\delta B / B)^2$, where $\delta B$ is the root-mean-squared value of the turbulent magnetic field. The typical scale of turbulent structures is described by the coherent length $l_{\rm cl} = \eta_{\rm cl} R_{\rm co} \lesssim R_{\rm co}$, where $\eta_{\rm cl}$ is the fractional scale.

Using $l_{\rm cl}$ and the diffusion coefficient in momentum space $D_p \sim 0.1 \sigma_{\rm tur} p^2 c / l_{\rm cl}$, derived from particle-in-cell simulations \cite{Comisso:2019frj} as a function of proton momentum $p$, we write the energy-independent proton acceleration time as
\begin{equation}
t_{\rm acc} \equiv \frac{p^2}{D_p} = \frac{10 \, l_{\rm cl}}{\sigma_{\rm tur} c}
\simeq 10^4 M_7 \left(\frac{\mathcal R}{20}\right) \left( \frac{\eta_{\rm cl}}{\sigma_{\rm tur}} \right) {\rm~s}.
\end{equation}
We then solve the time-dependent Fokker–Planck (FP) equations describing second-order Fermi acceleration, {which accounts for the feedback of cascade photons on proton cooling} initiated by $\gamma\gamma$ pair production and hadronic processes including photopion ($p\gamma$) interaction, proton-proton ($pp$) collision, Bethe–Heitler (BH) pair production, and proton synchrotron (p-sy) radiation. These processes are simulated using the open-source \am \cite{Klinger:2023zzv} software for time-dependent modeling of leptonic-hadronic interactions, where the cooling timescale can be retrieved as
$t_{\rm cool} = ( t_{p\gamma}^{-1} + t_{pp}^{-1} + t_{\rm BH}^{-1} + t_{ \rm p\rm-sy}^{-1} )^{-1}$,
and is used as an input to the FP equation. Regarding the proton escape timescale ($t_{\rm esc}$), the proton mean free path is determined by the coherent length of turbulent magnetic field fluctuations and the proton gyro-radius $l_r = E_p / (eB)$, where the proton energy is $E_p = pc$. This can be expressed as
$\lambda_{\rm mfp} = l_{\rm cl} (l_r / l_{\rm cl})^\zeta$,
with $\zeta > 0$. The exact value of $\zeta$ depends on the turbulence intermittency, and simulations suggest $0.3 < \zeta < 0.5$ \cite{Lemoine:2023sxw,Kempski:2023ikw,Gorbunov:2025ges}. We adopt $\zeta = 1/3$ as a fiducial value. The escape time, inspired by random walk arguments, is then
$t_{\rm esc} = R_{\rm co}^2 / (\lambda_{\rm mfp} c)$,
with a lower limit of $R_{\rm co}/c$. 

Given the acceleration time $t_{\rm acc}$, cooling time $t_{\rm cool}$, and escape time $t_{\rm esc}$, we use the numerical framework presented in \S\ref{sec:numerical} to model the proton acceleration and radiation processes.
The spectra of EM cascades and neutrinos are computed directly by \am, whereas the proton spectra are obtained from solving the FP equation. Importantly, we find that the accelerated proton spectra are largely insensitive to the injection term $q(p)=dN_p/(dtdVd^3p)$ in the momentum space, especially in the high-energy regime. Therefore, in applications to NGC~1068 and transient corona of TDE AT 2019dsg, we adopt
$q(p) \propto p \exp(-p/p_{\rm inj})$ with $p_{\rm inj} = 10 m_p c$, which yields an accurate proton distribution for $E_p > p_{\rm inj} c\sim10~\rm GeV$.

\subsection{Benchmark: steady corona in NGC 1068}\label{subsec:ngc1068}
The nearby Seyfert galaxy NGC~1068 (luminosity distance $d_L\simeq10.1$ Mpc \cite{Tully:2007ue}) provides an exceptional case study for probing proton acceleration and hadronic processes in a steady corona, as its $\sim 1$–$100$~TeV neutrino spectrum, combined with faint gamma-ray emission, favors a coronal origin. In this scenario, $p\gamma$ interactions with dense X-ray photons dominate neutrino production, while efficient $\gamma\gamma$ attenuation explains the low gamma-ray flux at $E_\gamma \gtrsim 0.1$~GeV. 

We apply the coupled time-dependent proton acceleration by magnetized turbulence to reproduce the neutrino spectrum and study the associated EM cascade. For a magnetically powered corona, the proton luminosity is constrained by the dissipation of turbulent magnetic energy. We approximate the dissipation timescale as the magnetic reconnection time scale $t_{\rm diss} \approx B l_{\rm cl}/(\epsilon_{\rm rec} v_A \delta B)$ \cite{Comisso:2019frj,Comisso:2024iyx}, where $v_A = c \sqrt{\sigma_B/(1+\sigma_B)}$ is the Alfv\'en speed and $\epsilon_{\rm rec} \sim 0.1$ defines the reconnection rate, e.g., $\epsilon_{\rm rec} v_A \delta B$ \cite{Comisso:2016ima,Cassak:2017enb}. The proton acceleration power is then parameterized as a fraction $\epsilon_p$ of the turbulent magnetic energy dissipation rate $L_B$ within the corona of volume $V_{\rm co} = 4\pi R_{\rm co}^3 / 3$,
\begin{equation}
    L_p = \epsilon_p L_B = \frac{\epsilon_p}{t_{\rm diss}} \frac{(\delta B)^2}{8\pi} V_{\rm co},
\end{equation}
which is used to normalize the accelerated proton spectra.
In the following calculations, we fix $\mathcal R=20$, $\sigma_B = \sigma_{\rm tur} = 0.1$, and assume $\delta B / B \sim 1$. In contrast, $\eta_{\rm cl}$ and $\epsilon_p$ are varied, e.g., $\epsilon_p,~\eta_{\rm cl}\leq1$, since they determine the maximum proton energy $E_{p, \rm max}$ through $t_{\rm acc}$ and the flux levels of neutrinos and EM cascades, respectively. {The magnetization $\sigma_B=0.1$ corresponds to the plasma beta $\beta=\sqrt{8\pi n_pk_BT/B^2}\sim20$ for a mildly relativistic temperature $k_BT\sim m_pc^2$.}

We consider isotropized OUV and coronal X-ray photons as $p\gamma$ targets. The external photon injection rates to \am, $\dot n_i(E_\gamma)\equiv E_\gamma d^2n_{i}/(dE_\gamma dt)$ with $i=$ OUV or X, are inferred from the luminosity distributions $dL_{i}/dE_\gamma$ via $\dot n_i=(dL_i/dE_\gamma)/V_i$, where $V_{\rm OUV}=4\pi R_{\rm OUV}^3/3$ and $V_X=V_{\rm co}$. OUV photons primarily originate from the accretion disk and the radius $R_{\rm OUV}\gtrsim10R_{\rm co}$ is estimated using the multi-temperature disk profile as 
\begin{equation}
    R_{\rm OUV}\approx\left(\frac{3GM L_{\rm OUV}}{8\pi\eta_{\rm rad}c^2\sigma_ST_d^4}\right)^{1/3},
\end{equation}
with $L_{\rm OUV}=\int dE_\gamma (dL_{\rm OUV}/dE_\gamma)$, radiation efficiency $\eta_{\rm rad}\sim0.1$, disk temperature $T_d$, and Stefan-Boltzmann constant $\sigma_S$. For NGC 1068, the blue curve in Fig.~\ref{fig:NGC1068} shows the OUV spectrum {(retrieved from Refs. \cite{Fiorillo:2024akm,Marconi:2003tg})} with peak temperature of $T_d\simeq4\times10^4$ K and $L_{X,\rm bol} = \int dE_\gamma \, (dL_X/dE_\gamma) \sim 10^{44}~\mathrm{erg\,s^{-1}}$ \cite{Fiorillo:2024akm}. We ignore the contribution from infrared photons \cite{Mullaney:2011iq}, since they are produced in a further extended region. {We use the coronal X-ray spectra from Ref. \cite{Murase:2019vdl}, normalized to $L_{\rm X,\rm bol}$ (green curve).} Proton cooling and hadronic emissions from BH pair productions, proton synchrotron, and $pp$ collisions are also included, with $n_p$ as the target proton density which is much larger than the density of accelerated protons.

With this configuration, the left panel of Fig.~\ref{fig:NGC1068} illustrates the coevolution of neutrino (dashed) and EM cascade (solid) spectra from $0.1t_{\rm fs}$ to $500t_{\rm fs}$, where $t_{\rm fs}=R_{\rm co}/c$ represents the free escaping timescale for photons and neutrinos. {The neutrino spectra and the proton spectra are consistent with previous work, such as Ref. \cite{Fiorillo:2024akm}.} Using $\epsilon_p=0.1$ and {$\eta_{\rm cl}=0.3$}, the neutrino and EM spectra converges efficiently to the steady state at $t\sim300t_{\rm fs}$ (equivalent to $\sim10t_{\rm acc}$), which reproduces the neutrino observations without exceeding the gamma-ray upper limits, since the
GeV gamma-rays observed by Fermi-LAT \cite{Fermi-LAT:2010feq}
would be produced in a different, extended region \cite{Lenain:2010kc,Yoast-Hull:2013qfa,Ambrosone:2021aaw, Eichmann:2022lxh,Lamastra:2016axo,Inoue:2022yak,Peretti:2023xqk,Yasuda:2024fvc,Dekker:2025vcg, Fermi-LAT:2025chw}. The early-time ($t\lesssim5t_{\rm fs}$) low-energy neutrino spectra are driven by $pp$ interactions of low-energy protons, whereas $pp$ and BH processes jointly dominate the EM cascades. 
As the maximum proton energy $E_{p,\rm max}$ increases (as shown in the right panel of Fig. \ref{fig:NGC1068}), $p\gamma$ interactions dominate neutrino production, and EM cascades generate broader EM spectra. In particular, coronal X-ray photons efficiently deplete gamma-rays above $10^7$ eV via $\gamma\gamma$ attenuation, with the resulting electron/positron ($e^\pm$) pairs reprocessing the energy into lower-energy radiation down to $\sim 0.01$ eV. The contributions of various radiation processes to the cascade spectra are shown in Appendix \ref{app:rates}. {The cascade spectra closely resemble those expected from reconnection scenarios (e.g., Refs. \cite{Fiorillo:2023dts,Karavola:2024uui}), making them ineffective discriminators of the proton acceleration mechanism in the corona.}

To verify the time-dependent treatment, we compare the time-dependent proton spectra (solid curves) with the steady-state solution (red dashed) obtained from the reduced FP equation in the right panel. Even with a low initial injection peak (e.g., $10$ GeV), protons are stably accelerated to $40$–$100$ TeV, consistent with the predicted value from balancing the acceleration and cooling rates. The deviation from the steady-state solution, which neglects cascade emissions, is negligible, as coronal X-ray photons dominate proton cooling.
\begin{figure*}[htp]\centering
    \includegraphics[width = 0.49\textwidth]{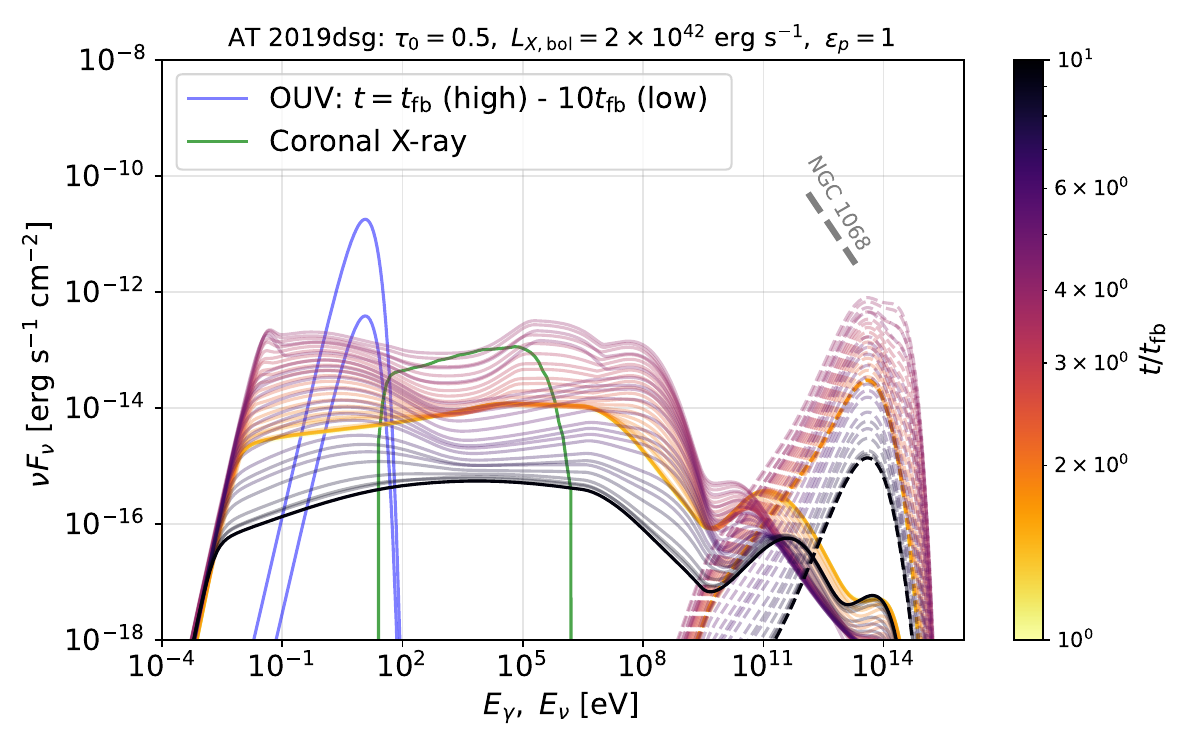}
      \includegraphics[width = 0.49\textwidth]{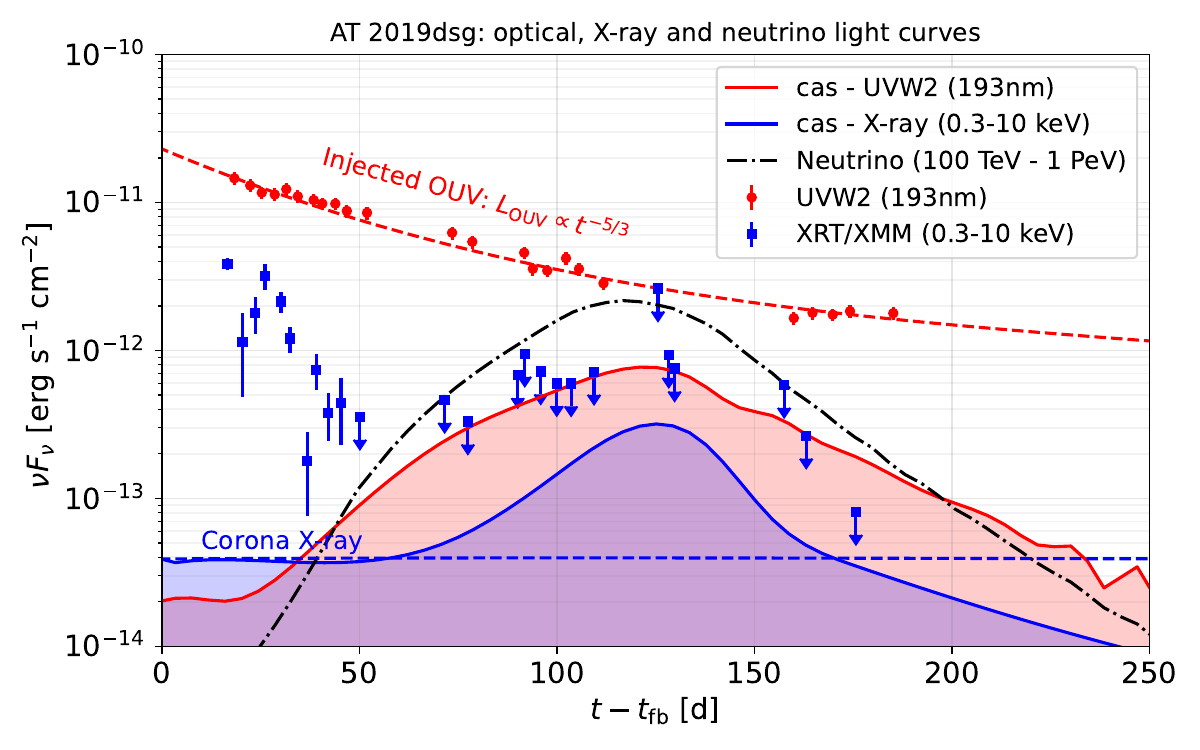}
        \caption{Left panel: Same as the left panel of Fig.~\ref{fig:NGC1068}, but for the transient corona of the TDE AT~2019dsg. The time is scaled to the TDE mass fallback time $t_{\rm fb}\simeq45$ d, up to $10t_{\rm fb}$. Right panel: Coronal cascade optical/UV (solid red), X-ray (solid blue), and neutrino (dash-dotted black) light curves of TDE AT 2019dsg. The injected OUV and X-ray light curves are shown as dashed lines. Red and blue points represent the UV (193 nm) \cite{Stein:2020xhk} and X-ray (0.3–10 keV) \cite{Cannizzaro:2020xzc} observations, respectively.}
    \label{fig:TDE_EM_NU}
\end{figure*}

\subsection{Extension to transient TDE coronae}\label{sec:app_TDE}

The previous application to NGC 1068 demonstrates the effectiveness of combining FP proton acceleration with radiation processes. We now take a step further to investigate the time-dependent signatures of TDEs. Although the dynamics and physical properties of SMBH coronae remain uncertain, we develop a phenomenological model for TDE coronae by {considering a transient corona} powered by a TDE with a characteristic mass fallback rate $\propto t^{-5/3}$. We adopt AT~2019dsg as a prototype, as it is a well-identified TDE with multi-wavelength light curve measurements in the radio, infrared, OUV, and X-ray bands \cite{Stein:2020xhk,vanVelzen:2021zsm}. Basic properties of AT~2019dsg are as follows \cite{Cannizzaro:2020xzc,Stein:2020xhk,vanVelzen:2021zsm}: redshift $z = 0.051$ (luminosity distance $d_L \simeq 228.3~\mathrm{Mpc}$), SMBH mass $M \sim 5 \times 10^{6}$–$10^{7} M_\odot$, and a peak bolometric OUV luminosity $L_{\rm OUV,pk} \simeq 10^{44}~\mathrm{erg~s^{-1}}$ characterized by a blackbody spectrum with temperature $k_B T_{\rm OUV} \sim 3.4~\mathrm{eV}$, where $k_B$ is the Boltzmann constant.

Using these parameters, the typical mass fallback time can be estimated as $t_{\rm fb}\simeq3.9\times10^6M_7(M_\star/M_\odot)^{-1/10}\rm ~s$, where $M_\star\sim0.5-2M_\odot$ \cite{Winter:2022fpf,Mohan:2021flu} is the mass of the disrupted star. After $t_{\rm fb}$, the OUV luminosity and mass accretion rate onto the SMBH for $t>t_{\rm fb}$ can be described respectively as $L_{\rm OUV}(t)=L_{\rm OUV,\rm pk}\left({t}/{t_{\rm fb}}\right)^{-5/3}$, as illustrated by the red dashed curve in Fig.~\ref{fig:TDE_EM_NU} (right panel), and
\begin{equation}
    ~\dot M=\frac{\eta_{\rm acc}M_\star}{3t_{\rm fb}c^2}\left(\frac{t}{t_{\rm fb}}\right)^{-5/3},
\end{equation}
where $\eta_{\rm acc}\sim0.01-0.1$ \cite{Murase:2020lnu,Yuan:2024daj,Yuan:2024sxk} denotes the fraction of gravitationally bound stellar mass that is ultimately accreted, and $\dot{M}$ is normalized by $\int \dot{M}\, dt = \eta_{\rm acc} M_\star / 2$, as roughly half of the disrupted star's mass remains bound.

Since the coronal protons may originate from accreted material, we take $n_p \propto \dot{M}$ as a reasonable approximation. Consequently, the Thomson opacity $\tau_T=n_p\sigma_T R_{\rm co}$ evolves as 
$\tau_T = \tau_{0}\min[1, \dot{M}/\dot{M}_{\rm Edd}]$, 
where $\dot{M}_{\rm Edd} = L_{\rm Edd}/(\eta_{\rm rad}c^2)$ is the Eddington accretion rate, $L_{\rm Edd} \simeq 1.3 \times 10^{46} M_7~\rm erg~s^{-1}$ is the Eddington luminosity, and $\tau_0 \sim 0.1$–$1$ represents the opacity in the super-Eddington phase. We keep $R_{\rm co}$ fixed, since the expansion of $R_{\rm co}$ would render the corona unstable and cause it to quickly merge with the extended disk winds. Given the same $\sigma_B = \sigma_{\rm tur} = 0.1$, $\tau_0=0.5$, and {$\eta_{\rm cl}=0.3$} as used for NGC~1068, we obtain the proton luminosity at $t_{\rm fb}$, $L_{p}(t_{\rm fb})=2.2\times10^{43}\epsilon_{p}~\rm erg~s^{-1}$. 

The X-ray emissions from TDE coronae are typically faint compared to NGC 1068. For AT~2019dsg, the X-ray observations \cite{Cannizzaro:2020xzc} (see the blue points in the right panel of Fig. \ref{fig:TDE_EM_NU}) imply a coronal luminosity of $L_{X,\rm bol} \lesssim 2 \times 10^{42}~\rm erg~s^{-1}$ in the 0.1–100~keV range (similar to Refs. \cite{Winter:2022fpf,Yuan:2023cmd}, where constant X-ray light curves are assumed). The initial fast decaying X-ray light curve could be caused by accretion disk cooling \cite{Cannizzaro:2020xzc}. In addition, thermal OUV photons produced at the blackbody radius 
\begin{equation}
R_{\rm OUV} \sim \sqrt{\frac{L_{\rm OUV}}{4 \pi \sigma_S T_{\rm OUV}^4}}
\end{equation}
also contribute to proton cooling. The blue and green curves in the left panel of Fig.~\ref{fig:TDE_EM_NU} show the spectra of the OUV {with fixed $T_{\rm OUV}$} and {AGN-like} coronal X-ray target photons, where $R_{\rm OUV}$ and $R_{\rm co}$ are used to infer the photon injection rates to \am. 

Applying the transient corona model, we evolve proton acceleration and radiation processes from $t_{\rm fb}$ to $10 t_{\rm fb}$. We adopt optimistic values $\eta_{\rm acc}=0.1$ and $\epsilon_p = 1$ to maximize coronal multiwavelength and neutrino emissions. The left panel of Fig.~\ref{fig:TDE_EM_NU} shows the resulting neutrino (dashed) and EM cascade (solid) spectra. The faint coronal X-ray emission leads to a higher $E_{p,\rm max}$ due to a lower $p\gamma$ cooling rate (see Appendix \ref{app:rates} for details), causing the neutrino spectrum to peak at $E_\nu \sim 100$ TeV. {The EM cascades dominate proton cooling above 1 PeV, and they also enhance the cooling rates in the 1 TeV $-$ 1 PeV range by a factor of $\sim2$}.

Interestingly, delayed OUV and X-ray cascade emissions are expected, as shown in the right panel of Fig.~\ref{fig:TDE_EM_NU}. This feature mainly arises from early-stage EM cascade feedback. In the $p\gamma$-efficient limit, where $f_{p\gamma}=t_{p\gamma}^{-1}/(t_{\rm cool}^{-1}+t_{\rm esc}^{-1}) \to 1$, the initial cascade luminosity $L_{\rm cas} \sim (5/8)L_p$ exceeds the coronal X-ray luminosity. The buildup of EM cascades, together with the accumulation of protons during the super-Eddington phase and for $t < t_{\rm esc}$, jointly drive the multiwavelength light curves to peak at $t \simeq 140$~d, consistent with the spectra in the left panel of Fig. \ref{fig:TDE_EM_NU} and with the OUV (193 nm, red solid) and X-ray (0.3–10 keV, blue solid) light curves in the right panel.

A similar interpretation applies to the neutrino light curve (0.1–1 PeV, black dash-dotted). The coronal neutrino peak flux and its time delay relative to the OUV peak ($\sim140$ d) are comparable to those of the isotropic radiation zone extending to the dust torus \cite{Winter:2022fpf,Yuan:2023cmd}, but the coronal neutrinos peak at lower energies ($\sim100$ TeV), offering a new perspective for interpreting the potential neutrino–TDE correlations \cite{Stein:2020xhk,Reusch:2021ztx,vanVelzen:2021zsm,Jiang:2023kbb,Yuan:2024foi,Li:2024qcp}, especially the time delays of $\mathcal O(100~\rm d)$. {For a decaying $L_{X,\rm bol} \propto t^{-5/3}$, $E_{p,\rm max}$ would increase over time and the light curves are expected to decline faster after the peak, following $f_{p\gamma}L_p \propto L_{X,\rm bol} L_p\propto t^{-10/3}$.} One caveat is that the OUV, X-ray and neutrino cascade light curves in the right panel represent upper limits for AT 2019dsg, since in reality—especially in the weak corona case—lower values of $\tau_0\lesssim0.5$ and $\epsilon_p\lesssim1$ are expected. In such cases, a correction factor of $\sim2\tau_0 \epsilon_p$ should be applied to the neutrino and EM cascade fluxes.

\begin{figure*}[htp]
    \centering
    \includegraphics[width=0.49\linewidth]{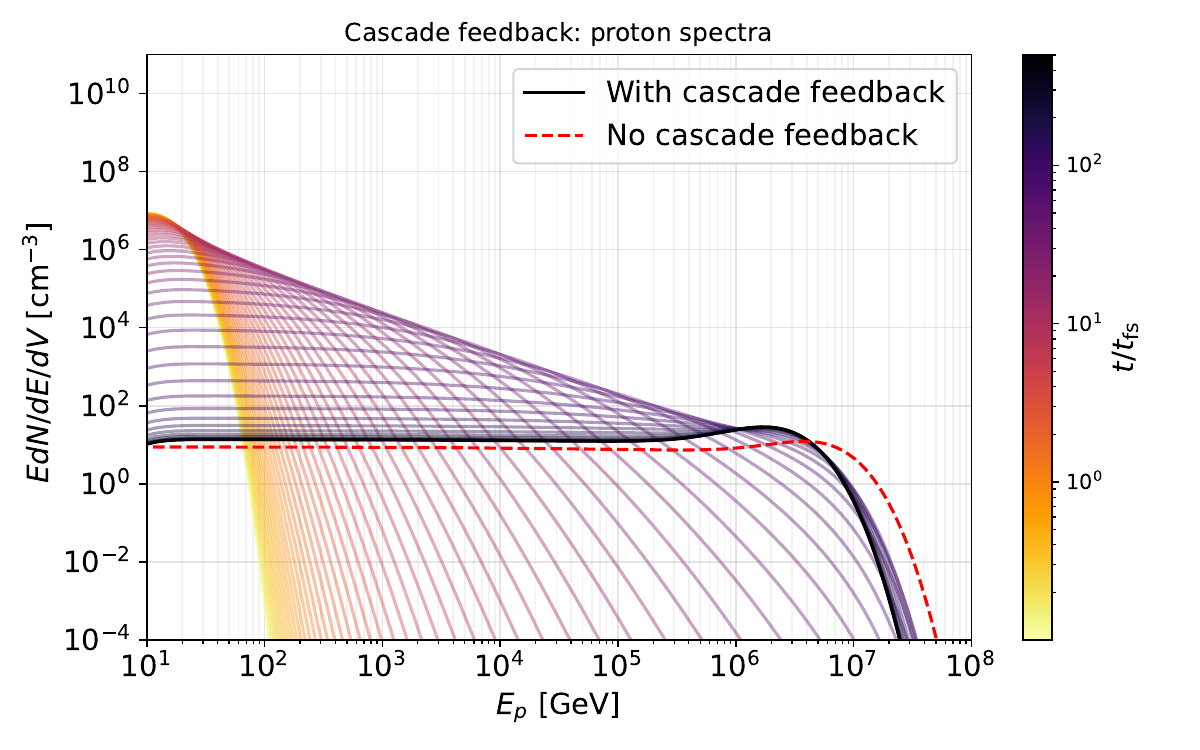}
    \includegraphics[width = 0.49\linewidth]{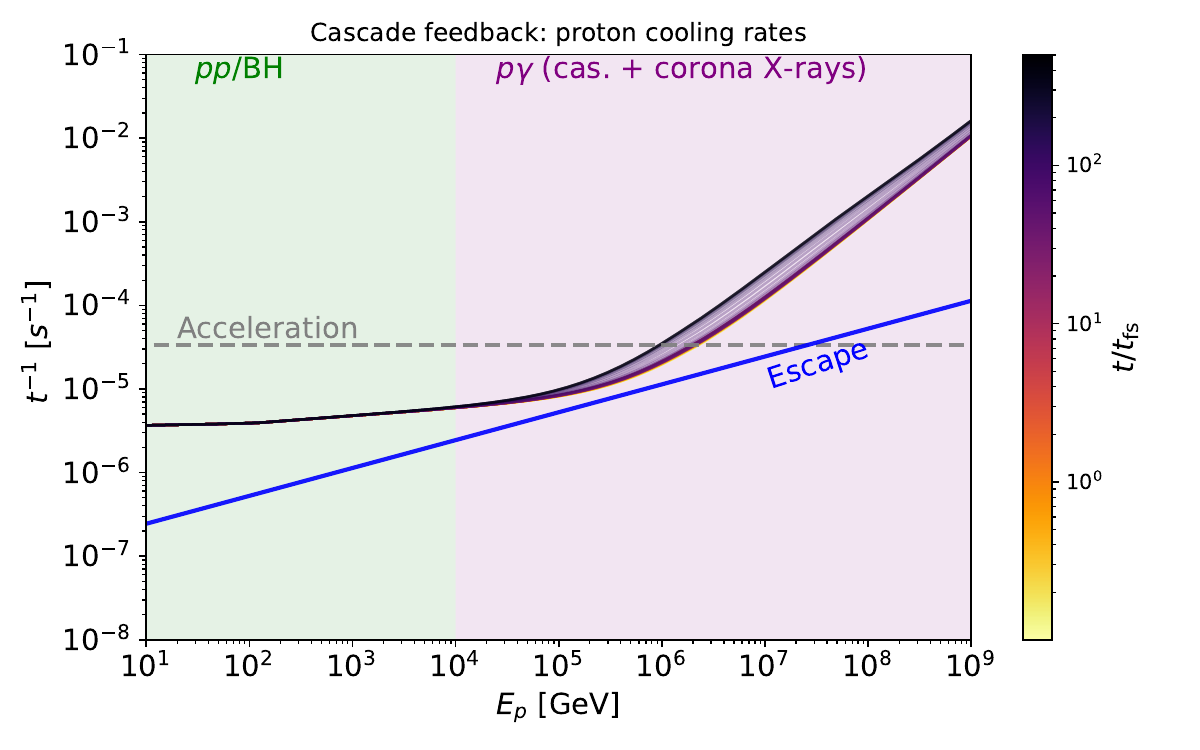}
    \caption{Test of the impact of the EM cascade feedback for a corona with weaker X-ray luminosity (e.g., $L_{X,\rm bol} < L_p$).
NGC 1068's corona is used as an example, with parameters identical to those used in \S\ref{subsec:ngc1068} except for a lower luminosity of $L_{X,\rm bol} = 5\times10^{41}\rm erg~s^{-1}$. Left panel: Proton spectra obtained with EM cascade feedback (solid curves) and without cascade feedback (red solid curve). Right panel: Proton cooling rates at times from $0.1t_{\rm fs} = 0.1R_{\rm co}/c$ to $500t_{\rm fs}$, including the contribution from cascade photons. The acceleration and escape rates are also shown. The green and purple areas represent the regimes where $pp$/BH and $p\gamma$ processes, respectively, dominate proton cooling.}
    \label{fig:Sup_test_feedback}
\end{figure*}

\section{Discussion}\label{sec:discussion}
\subsection{Radiation feedback}
We quantify the impact of radiation feedback on the particle-acceleration process and validate our findings numerically. In complex scenarios, particularly when 
$L_p$ is comparable to or higher than the injection power of external photons, feedback to the radiation field via electromagnetic cascades induced by hadronic processes increases the cooling rates. The resulting additional photons enhance the $p\gamma$ and BH interactions, which can reduce the peak energy of accelerated protons. 

{Fig. \ref{fig:Sup_test_feedback} shows the proton spectra (left panel) and cooling rates (right panel) for this case, using NGC 1068 as an example. All physical setups and parameters are the same as in the main paper, except that a lower coronal X-ray luminosity of $L_{X,\rm bol} = 5\times10^{41}~\mathrm{erg~s^{-1}} < L_p\simeq4\times10^{42}~\rm erg~s^{-1}$ is used. The results demonstrate that the EM cascade could enhance the proton cooling rate by a factor of 
\[f_{\rm cas}\sim\max\left[1,~\frac{5f_{p\gamma}L_p}{8\mathcal CL_{X,\rm bol}}\right],\]
where $f_{p\gamma}=t_{p\gamma}^{-1}/(t_{\rm cool}^{-1}+t_{\rm esc}^{-1})$ is the $p\gamma$ efficiency at $E_{p,\rm max}$ and the correction factor $\mathcal{C} \sim 2\text{–}3$ accounts for the difference in the widths of the cascade and coronal X-ray spectra. Consequently, the proton maximum energy shifts to lower energies.}

{\subsection{Turbulent particle acceleration}\label{sec:discussion_acc}
{The momentum-dependent diffusion coefficient, cooling time, and escape time are used as inputs to the FP equations to model proton acceleration in turbulent coronae. In particular, the acceleration and escape times are taken from existing particle-acceleration studies, including theoretical models and PIC simulations. Recent progress has shown that additional modifications are often required to make the current FP treatment consistent with simulation results and applicable to more general cases. Although this work aims to establish a numerical framework that links particle-acceleration physics to time-dependent multi-messenger emissions, we outline several directions for improving the proton acceleration model in future studies.

\emph{Momentum diffusion coefficient $D_{p}$.} In this study, the acceleration time scale, $t_{\rm acc}=p^2/D_p$, determines the low-energy proton spectrum, since $t_{\rm acc}<\min(t_{\rm cool},~t_{\rm esc})$ for $E_p\lesssim10^4$ GeV (see Fig.~\ref{fig:Sup_spectra_rates}). Refs.~\cite{Wong:2019dog,Wong:2025kzf} have demonstrated that $D_p$ can be divided into two segments: a universal scaling $D_p\propto p^2$ at high energies and a weaker momentum dependence $D_p\propto p^{\alpha}$ with $\alpha\sim2/3-1$ at low energies, depending on the magnetization. This modification primarily affects the low-energy proton spectrum, which ensures that the fitting to the neutrino spectrum in Fig.~\ref{fig:NGC1068} remains valid (as it depends mainly on the spectral shape near the peak energy).

\emph{Advection.} It has been found that an additional advection term $A(p)$ is typically required in a diffusive FP equation (Eq. \ref{eq:FPT1}) to reproduce the accelerated particle spectra obtained from PIC simulations \cite[e.g.,][]{Wong:2019dog,Lemoine:2023wsw}. This additional term has a weak dependence on momentum (e.g., $x-y\ln[1+p/(m_pc)]$ as given by Ref. \cite{Wong:2019dog}, where $x$ and $y$ are positive coefficients), and consequently narrows the particle distribution by accelerating low-energy particles and decelerating high-energy particles. In practice, to account for this effect, one may absorb $A(p)$ into the radiation cooling by applying the following correction to Eq. \ref{eq:FPT1}:
\begin{equation}
\frac{p}{t_{\rm cool}}\rightarrow \frac{p}{t_{\rm cool}} - A(p),
\end{equation}
since $p/t_{\rm cool}$ and $A(p)$ respectively describe the cooling-induced and noise-induced drifts. In the corona scenario, cooling is highly efficient and controls the location of the proton spectral peak; however, in cases where cooling is weak, the advection effect could significantly alter the peak energy of the proton distribution.

\emph{Proton acceleration feedback on turbulence.} From the proton spectra in Fig. \ref{fig:NGC1068}, we estimate the fraction of accelerated protons drawn from the thermal reservoir to be $\sim10^{-5}$, which implies that the energy carried by the accelerated protons is comparable to the pressure of the background plasma, assuming a mildly relativistic temperature $k_BT\sim m_pc^2$. Ref. \cite{Lemoine:2023sxw} showed that this fine-tuning coincidence can be interpreted as a natural consequence of feedback on the turbulence cascade by the accelerating protons, which self-regulates the acceleration \cite{Lemoine:2024roa}. Studying the feedback exerted by accelerating protons constitutes a separate problem and is not intended to be addressed in this paper.

\emph{Additional corrections.} Motivated by simulations \cite[e.g., Refs.][]{Lemoine:2023sxw,Kempski:2023ikw,Gorbunov:2025ges}, the escape time $t_{\rm esc}\propto p^{-1/3}$ is used. This scaling can cause low-energy particles to be strongly confined, resulting in a hard proton spectrum. Recent ultra-high-resolution MHD simulations suggest that the escape time may have a much weaker energy dependence \cite{Kempski:2025goy}. Moreover, additional mechanisms can modify the accelerated proton spectra, such as (a) particle streaming along tangled magnetic field lines for $\delta B\sim B$ turbulence; (b) intermittent acceleration \citep[e.g., formulated by Ref.][]{Lemoine:2024roa}; and (c) the increasing enthalpy per particle and the decreasing magnetization due to proton heating, both of which slow down acceleration. A systematic investigation of how these effects alter the results would be valuable, but is beyond the scope of this paper.

{\subsection{Corona model}}
\emph{Coronal X-rays.} We treated the coronal X-ray emission as an external photon field characterized by the SMBH parameters and the X-ray luminosity $L_{X,\rm bol}$ as in Ref. \cite{Murase:2019vdl}, and do not attempt to model its origin due to the high complexity of the problem and the fact that the underlying mechanism remains under debate. It is widely accepted that hard X-rays are produced through inverse Compton scattering of UV and soft X-ray photons originating from the accretion disk \cite[e.g.,][]{Mayer:2006vf,Maraschi:1996ey}. However, the electrons responsible for this emission may arise from pair production and/or be supplied by the disk through magnetic buoyancy or thermal evaporation. These electrons can be energized by plasma heating and various particle-acceleration processes \cite[e.g.,][]{Sridhar:2021bvf}, producing a high-energy tail extending to the MeV range (in addition to the gamma-ray emission from EM cascades) \cite[e.g.,][]{Wardzinski:2001fj,Malzac:2001av}. The relevant processes are highly nonlinear, as feedback on the turbulence cascade from accelerating protons could strongly suppress electron (and positron) acceleration \cite{Lemoine:2024roa} and thus the resulting coronal X-rays. Incorporating particle-acceleration feedback into the current acceleration–radiation framework may ultimately help clarify this issue, but doing so would require substantial additional effort.

\emph{TDE transient coronae.} In \S\ref{sec:app_TDE}, we postulated a transient TDE corona and investigated the time-dependent acceleration and radiation processes. Although there is no \emph{direct} evidence for the existence of coronae in TDEs so far, we presented testable predictions. Recently, Ref. \cite{Li:2024qcp} reported a potential coincidence between a neutrino flare and a TDE X-ray flare: a bright neutrino flare accompanied by faint gamma-ray emission, analogous to the case of NGC 1068. They pointed out that isotropic wind models cannot simultaneously accommodate the high neutrino flux and satisfy the $\gamma$-ray constraints, which may hint at a coronal origin.}

}
\section{Summary}\label{sec:summary}
We have developed an efficient and stable numerical code to solve the Fokker-Planck equations describing proton accelerations and combined proton acceleration due to magnetized turbulence in coronae with leptonic-hadronic radiation modeling in a fully time-dependent manner. This framework has been applied to model the neutrino and EM cascade spectra from a steady-state and a transient/dynamic corona, motivated by NGC 1068 and TDEs. In the former case, the power densities of accelerated protons and EM cascades are lower than those of the injected X-ray and OUV target photon fields. {We find that radiation feedback onto the accelerated proton distribution is negligible, and the solution converges perfectly to a steady state that reproduces the neutrino spectra while being consistent with gamma-ray observations and upper limits.} For TDEs, where {the X-ray emission from the corona is typically weak}, EM cascade feedback can be more important. We propose a transient corona scenario for AT 2019dsg, predicting delayed OUV, X-ray, and neutrino emissions from early-stage EM cascade feedback.

The steady corona model can be directly applied to other neutrino-emitting Seyfert galaxies, such as NGC 4151, NGC 3079 \cite{Neronov:2023aks}, NGC 7469 \cite{Sommani:2024sbp}, and the Circinus galaxy \cite{Murase:2023ccp}, whereas the transient corona model is testable via multi-messenger observations of TDEs, especially for subpopulations with potential neutrino correlations or strong non-jetted X-ray emissions, as the coronal contribution would be prominent.

Beyond turbulent acceleration and SMBH coronae, this coupled acceleration–cascade framework has broader applications. Its flexible timescales and injection terms enable modeling of other mechanisms such as magnetic reconnection, shear-flow acceleration, and shock acceleration, {while allowing us to test how new results in particle-acceleration physics (see \S\ref{sec:discussion_acc}) imprint on the multi-messenger modeling.} The time-dependent feature suits both steady sources and transient multi-messenger phenomena, e.g., gamma-ray bursts, TDEs, and AGN flares, where proton acceleration could coexist with non-thermal radiation. Timely for IceCube-Gen2 \cite{IceCube-Gen2:2020qha} and KM3NeT \cite{KM3Net:2016zxf}, this framework bridges plasma dynamics and acceleration microphysics—addressed respectively by magnetohydrodynamic simulations (e.g., Refs. \cite{Lemoine:2021mtv,Bresci:2022awc,Price:2017mwk}) and particle-in-cell simulations (e.g., Refs. \cite{Comisso:2019frj,Wong:2019dog,Zhdankin:2019dfz,Bresci:2022awc})—with radiation modeling to improve the interpretation and prediction of high-energy astrophysical phenomena.

\section*{Acknowledgments}This work has been partially funded by the ``Program for the Promotion of Exchanges and Scientific Collaboration between Greece and Germany IKYDA--DAAD'' 2024 (IKY project ID 309; DAAD project ID: 57729829). M.P. acknowledges support from the Hellenic Foundation for Research and Innovation (H.F.R.I.) under the ``2nd call for H.F.R.I. Research Projects to support Faculty members and Researchers'' through the project UNTRAPHOB (Project ID 3013). Q.L. is supported by the Arthur B. McDonald Canadian Astroparticle Physics Research Institute. Research at Perimeter Institute is supported by the Government of Canada through the Department of Innovation, Science, and Economic Development, and by the
Province of Ontario. 

\bibliographystyle{apsrev4-2}

\bibliography{ref}

\appendix
\section{Numerical methods for solving the time-dependent and steady-state Fokker-Planck equations}\label{app:numerical}

We present a stable numerical method for solving the time-dependent Fokker-Planck (FP) equation that describes proton acceleration in momentum space (e.g., Refs. \cite{1970JCoPh...6....1C,park1996stochastic,Becker:2006nz,Stawarz:2008sp}),
\begin{equation}
    \frac{\partial f}{\partial t}=\frac{1}{p^2}\frac{\partial}{\partial p}\left[p^2D_{p}(p)\frac{\partial f}{\partial p}+\frac{p^3}{t_{\rm cool}} f\right] -\frac{f}{t_{\rm esc}}+q(p),
    \label{eq:FPT}
\end{equation}
where $f(p,t)=\frac{dN_p}{dp^3dV}$ is particle distribution, 
$q(p)=\frac{dN_p}{dp^3dVdt}$ is the proton injection rate, $t_{\rm cool}$ is the proton cooling timescale, and $t_{\rm esc}$ is the proton escape time scale.
The momentum diffusion term \( D_p \) is related to the acceleration timescale as \( D_p = p^{2} / t_{\rm acc} \). To solve the partial differential equation (PDE), we adopt linear and logarithmic discretizations for the time \( t \) and the momentum \( p \), respectively, with \( t_n = t_0 + n \Delta t \) (for \( n = 1, 2, \ldots \)) and \( p_i = p_0 \left( p_{\rm max} / p_0 \right)^{i/N} \) (for \( i = 0, 1, \ldots, N \)). The minimum and maximum proton momenta are set to \( p_0 = 5 m_p c \) and \( p_{\rm max} = 10^{11} m_p c \) to meet the requirements of most cases. A proton momentum grid with \( N = 274 \), corresponding to 25 bins per energy decade, is used to improve numerical stability. To simplify the presentation below, we define \( \dot{p} \equiv p / t_{\rm cool} > 0 \) as the momentum advection term and adopt the following notations
\begin{itemize}
    \item  $f_i^n \equiv f(p_i,t_n),~D_i\equiv D_{p}(p_i),~\dot p_i\equiv\dot p (p_i),~q_i\equiv q(p_i),\lambda_i \equiv 1/t_{\rm esc}(p_i)$
    \item  $x_{i+1/2}=(x_i+x_{i+1})/2,~\Delta x_{i+1/2}=x_{i+1}-x_i,~\Delta x_i = (x_{i+1}-x_{i-1})/2$ for $x=D_p,~p$ and $\dot p$. For instance, $p_{i+1/2}=(p_i+p_{i+1})/2$.
\end{itemize}
The flux term $\Phi(p,t)\equiv D_{p}(p)\frac{\partial f}{\partial p}+\dot p f$
could be discretized as 
\[\Phi^n_{i+1/2}=D_{i+1/2}\frac{f_{i+1}^n-f^n_i}{\Delta p_{i+1/2}}+\dot p_{i+1/2}[(1-\delta) f_{i+1}^n+\delta f_i^n],\]
where $0\leq\delta\leq1$ is the Chang-Cooper weighting factor \cite{1970JCoPh...6....1C} defined as
\[\delta=\frac{1}{w}-\frac{1}{e^w-1}\]
with $w=\Delta p_{i+1/2}\dot p_{i+1/2}/D_{i+1/2}$. This method ensures both the preservation of positivity and the correct equilibrium solution, even in strongly cooling-dominated ($w\gg1$) regimes. In practice, we find that \( w \gg 1 \), especially at high \( p \), where proton cooling dominates the spectral evolution and the resulting \( \delta \rightarrow 0 \), corresponding to the upwinding scheme. The physical meaning is that when the cooling rate is exceedingly high, protons flow from high \( p_{i+1} \) to low \( p_i \).

We then have the expression for $\partial f/\partial t$ at $p_i$ and $t_n$,
\[\frac{\partial{f}}{\partial t}|_{i,n}\approx \frac{1}{p^2_i\Delta p_i}\left[p_{i+1/2}^2\Phi^n_{i+1/2}-p_{i-1/2}^2\Phi^n_{i-1/2}\right]-{f_i^n}\lambda_i+q_i.\]
The time evolution of \( f_i \) is then computed using the Crank–Nicolson time discretization scheme \cite{CrankNicolson1947}, e.g.,
\[\frac{f_i^{n+1}-f_i^{n}}{\Delta t} = \frac{1}{2}\left[\frac{\partial f}{\partial t}|_{i,n+1}+\frac{\partial f}{\partial t}|_{i,n}\right].\] This implicit scheme is stable for linear problems while maintaining second-order accuracy in time, and it can be expressed in a tridiagonal matrix form as
\[A_if_{i-1}^{n+1}+B_if_i^{n+1}+C_if_{i+1}^{n+1}=R_i^n.\]
Defining the coefficients
\begin{equation}
    \begin{split}
        a_i &= \frac{p_{i+1/2}^2D_{i+1/2}}{p_i^2\Delta p_i\Delta p_{i+1/2}},~
        b_i = \frac{p_{i+1/2}^2\dot p_{i+1/2}}{p_i^2\Delta p_i},\\
        c_i & = \frac{p_{i-1/2}^2D_{i-1/2}}{p_i^2\Delta p_i\Delta p_{i-1/2}}
        ,~d_i  = \frac{p_{i-1/2}^2\dot p_{i-1/2}}{p_i^2\Delta p_i},\\
    \end{split}
\end{equation}
we explicitly write down
\begin{equation*}
\begin{split}
        A_i &= -\frac{\Delta t}{2}\left(c_i-\delta d_i\right),\\       
        B_i & = 1+\frac{\Delta t}{2}\left[a_i-\delta b_i+c_i+(1-\delta) d_i+\lambda_i\right],\\      
        C_i & = -\frac{\Delta t}{2}\left[a_i+(1-\delta) b_i\right],\\ 
        R_i^n & =\left(1-\frac{\lambda_i\Delta t}{2}\right)f_i^n+\frac{\Delta t}{2}\{a_i(f_{i+1}^n-f_i^n)\\
        &+b_i[(1-\delta) f_{i+1}^n+ \delta f_i^n]-c_i(f_i^n-f_{i-1}^n)\\
        &-d_i[(1-\delta) f_i^n+\delta f_{i-1}^n]+2q_i\}.\\
\end{split}
\end{equation*}
The system can efficiently evolve from  \( t_n \) to \( t_{n+1} \) by applying the inverse of a tridiagonal matrix \( \mathcal{M} \),
\begin{equation}
({\vec f_i}^{n+1})^{T}=\mathcal M^{-1}\cdot (\vec R^n_i)^{T},
\label{eq:FP_evolve}
\end{equation}
where $\vec f_i^{n+1}=(f_0^{n+1},...,f_N^{n+1})$, $\vec R_i^{n}=(R_0^{n},...,R_N^{n})$, and $\mathcal M$ can be explicitly written as
\begin{equation}
    \mathcal M = \begin{bmatrix}
B_0 & C_0 & 0        & 0        & \cdots & 0 \\
A_1 & B_1 & C_1 & 0        & \cdots & 0 \\
\vdots   & \ddots  & \ddots  & \ddots  & \ddots & \vdots \\
0        & \cdots  & 0       & A_{N-1} & B_{N-1} & C_{N-1} \\
0        & \cdots  & 0       & 0        & A_{N} & B_{N}
\end{bmatrix}.
\end{equation}
By repeating the above procedure \( K \) times and updating the input terms accordingly, the proton distribution at time \( t_K = t_0 + K \Delta t \) can be obtained.

{\it Steady-state solution.} To test the stability and convergence of the solutions to the time-dependent FP equation, it is useful to solve Eq.~\ref{eq:FPT} in the steady state by setting \(\partial f / \partial t = 0\). The PDE then reduces to an ordinary differential equation (ODE), which can also be expressed in tridiagonal matrix form, for example,
\[
\alpha_i f_{i-1} + \beta_i f_i + \gamma_i f_{i+1} = r_i,
\]
with the coefficients
\begin{equation*}
    \begin{split}
    &\alpha_i = c_i-\delta d_i,\\
    &\beta_i = -a_i +\delta b_i-c_i-(1-\delta)d_i-\lambda_i,\\
    &\gamma_i = a_i+(1-\delta)b_i, \\
    &r_i =-q_i.\\
\end{split}
\end{equation*} 
We impose a zero-flux boundary condition at \( p_0 \), i.e., \( \Phi(p_0) = 0 \), and an absorption boundary condition at \( p_{\rm max} \), i.e., \( f(p_N) = 0 \):
\[
D_{1/2} \frac{f_1 - f_0}{\Delta p_{1/2}} + \dot{p}_{1/2} \big[(1-\delta) f_1 + \delta f_0 \big] = 0, \quad f_N = 0.
\]
While solving this ODE, we found that the Chang–Cooper method remains effective in controlling the momentum flow direction.

\begin{figure*}
    \centering
    \includegraphics[width=0.49\linewidth]{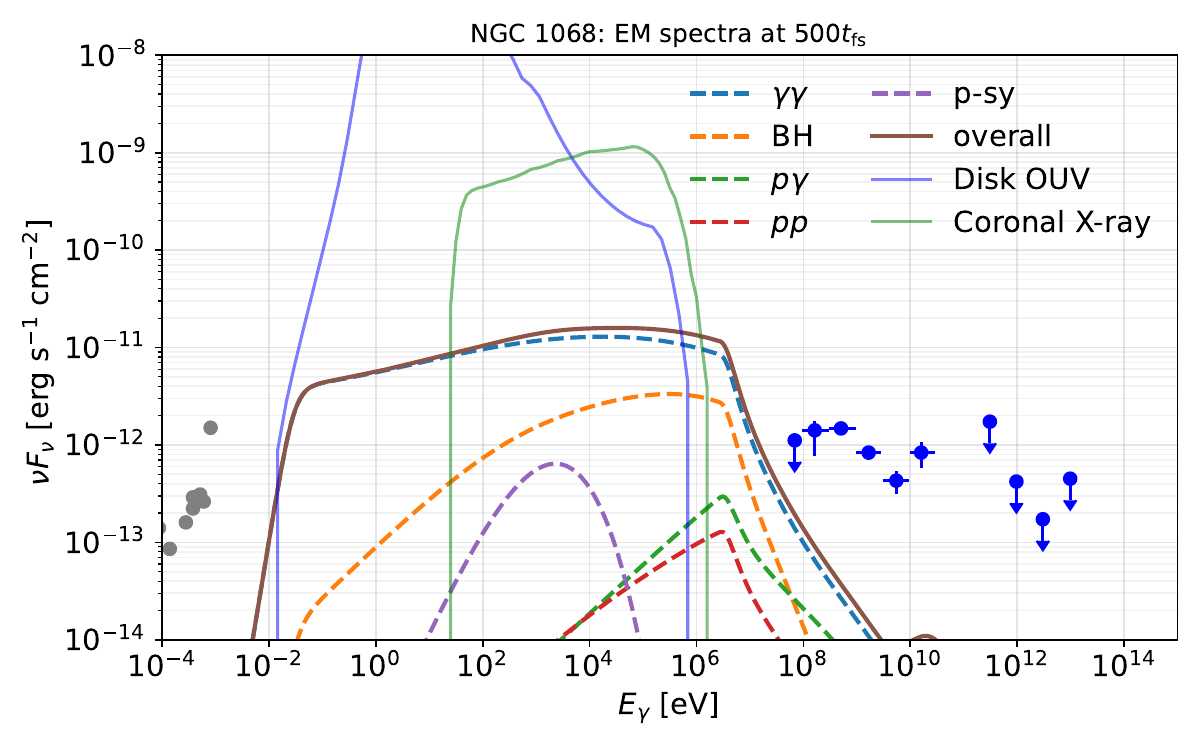}
      \includegraphics[width=0.49\linewidth]{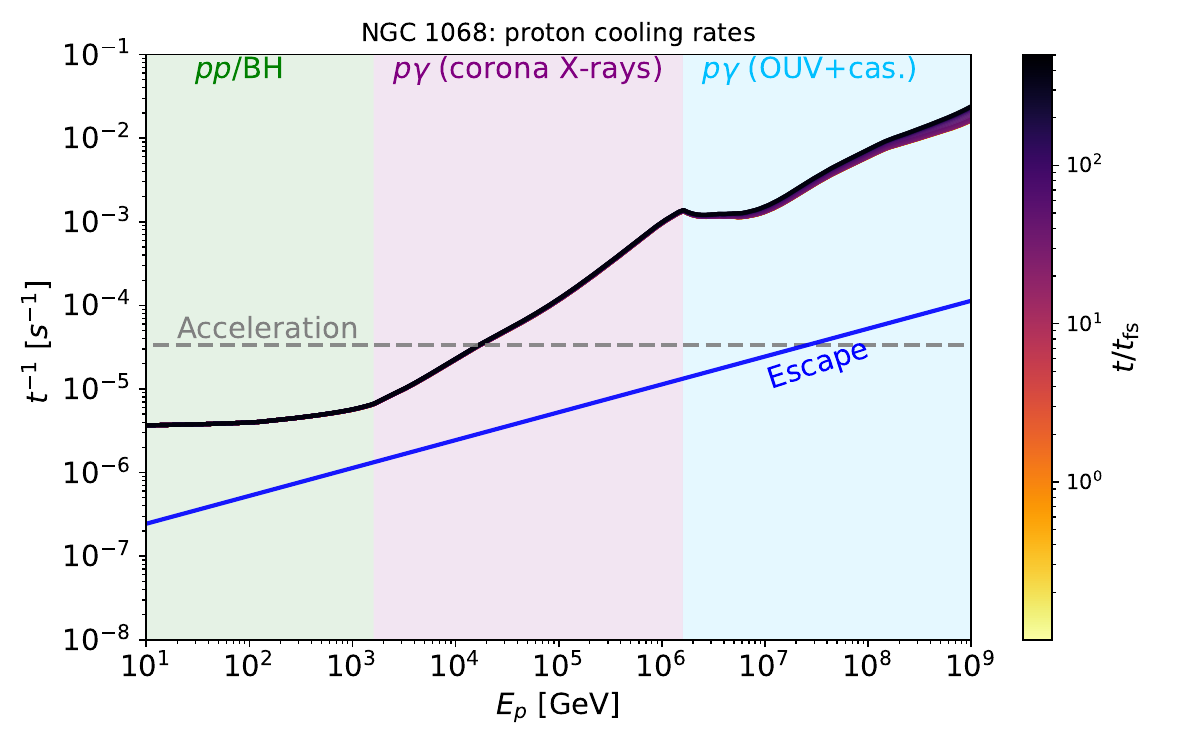}
      \includegraphics[width=0.49\textwidth]{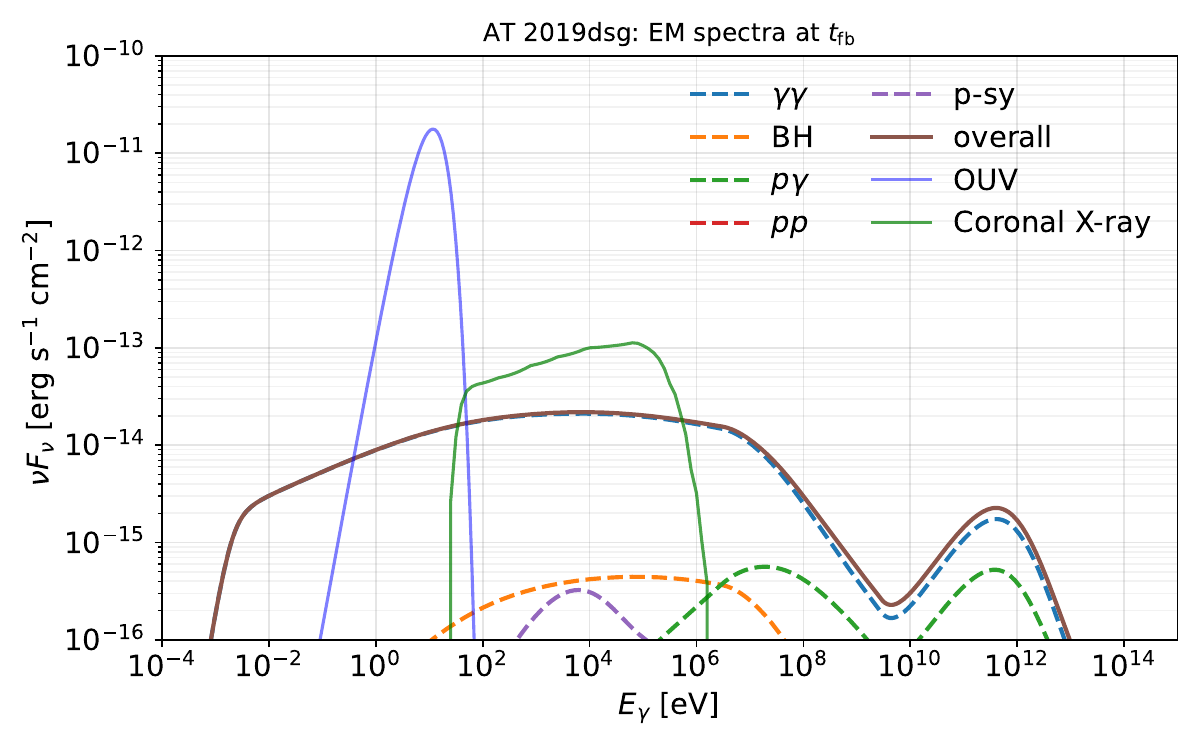}
    \includegraphics[width = 0.49\textwidth]{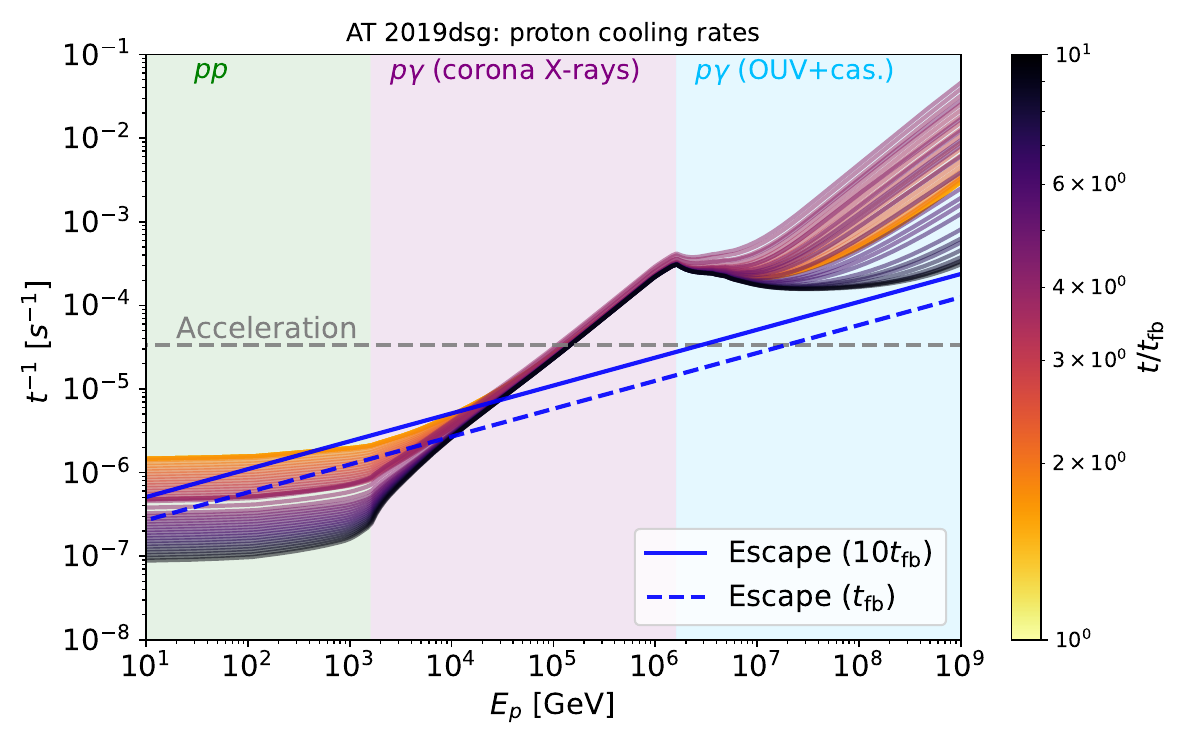}
    \caption{Left panels: EM cascade spectral components, including contributions from $\gamma\gamma$/BH pairs, leptons from $p\gamma$/$pp$ interactions, and proton synchrotron radiation, are shown for NGC 1068 (upper panel) at $500t_{\rm fs}$ and for the TDE AT~2019dsg (bottom panel) at $t_{\rm fb}$, where $t_{\rm fs}=R_{\rm co}/c$ is the free escaping time and $t_{\rm fb}$ is the TDE mass fallback time. Radio \cite{Chang:2019cdu} and gamma-ray \cite{Fermi-LAT:2019yla,MAGIC:2019fvw} observations of NGC 1068 are shown as the gray and blue points. The OUV and coronal X-ray spectra are also shown. Both calculations account for $\gamma\gamma$ attenuation by the extragalactic background light. Right panels: Proton cooling rates at various times are shown, with colors ranging from yellow to black. The horizontal dashed gray curve represents the acceleration rate, while the blue lines depict the proton escape rates. The shaded regions from left to right indicate proton cooling dominated by $pp$/BH interactions, by $p\gamma$ interactions with coronal X-ray photons, and by interactions with OUV and cascade photons.}
    \label{fig:Sup_spectra_rates}
\end{figure*}

\begin{figure}
    \centering
    \includegraphics[width=0.5\textwidth]{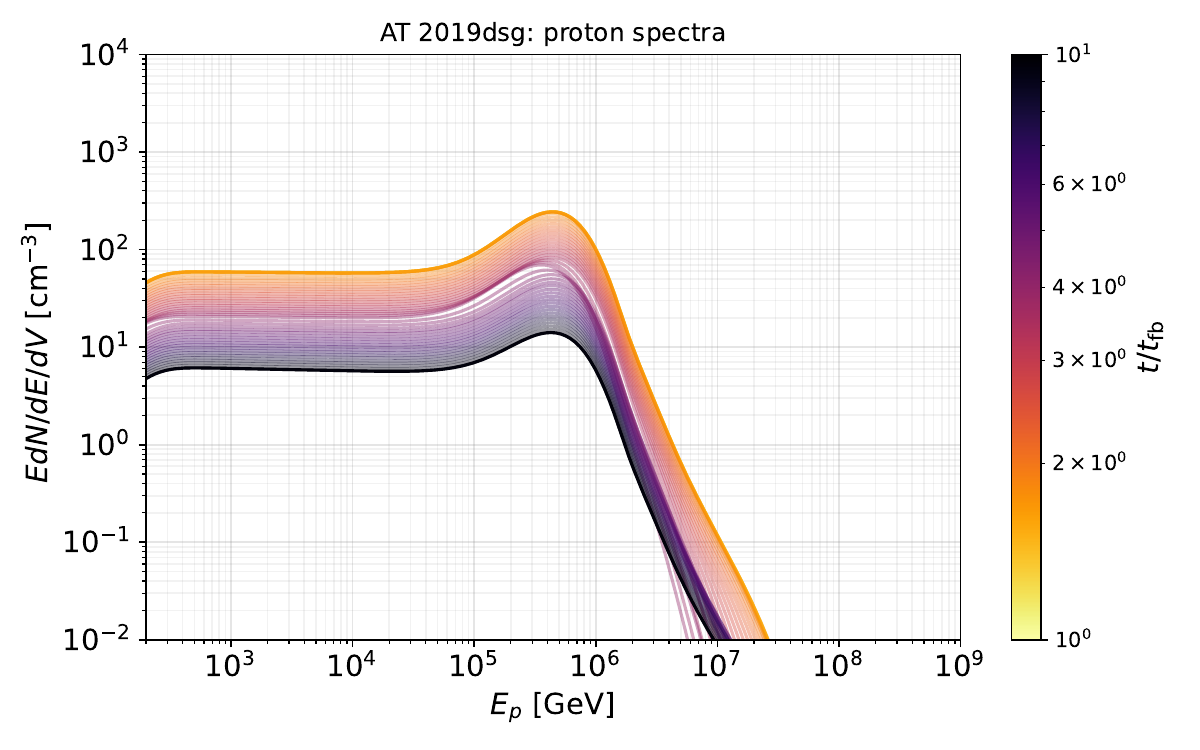}
    \caption{TDE AT 2019dsg: In-source proton density spectra at times ranging from $t_{\rm fb}$ to $10t_{\rm fb}$.}
    \label{fig:Sup_TDE_p}
\end{figure}

\section{EM {cascade spectra, proton cooling rates and} TDE {proton spectra}}\label{app:rates}
Here, we present the components of the electromagnetic (EM) cascade spectra and the proton cooling rates for NGC 1068 (upper panels of Fig. \ref{fig:Sup_spectra_rates}) and the tidal disruption event (TDE) AT 2019dsg (bottom panels of Fig. \ref{fig:Sup_spectra_rates}) to support the interpretation of the EM cascade, neutrino, and proton spectra discussed in the main text. The time evolution of proton spectra in TDE corona is also shown in Fig. \ref{fig:Sup_TDE_p}. The main conclusions drawn from these figures are summarized below:
\begin{itemize}
    \item In compact coronae, the late-stage EM spectra are universally dominated by $e^\pm$ pairs produced via $\gamma\gamma$ attenuation (see e.g., Refs. \cite{2025arXiv250900152F,Berezinsky:1975zz}), as the protons are accelerated to high energies, while in the early stage, EM cascades from $pp$ and BH processes could dominate (e.g., NGC 1068).
    
    \item The maximum proton energy approaches the value determined by $t_{\rm acc}^{-1} = t_{\rm cool}^{-1}$. 
    
    \item For AT 2019dsg, the buildup of EM cascades and accelerated protons during the super-Eddington phase (i.e., $\dot{M} > \dot{M}_{\rm Edd}$) and for $t < t_{\rm esc}$ jointly drive the peaks of the multiwavelength and neutrino light curves. 
\end{itemize}

\end{document}